\documentclass[aps,prd,tightenlines]{revtex4}

\usepackage{plain}
\usepackage{latexsym}
\usepackage{amsmath,amsthm,amsfonts,amssymb,bbm}
\usepackage[mathscr]{eucal}
\usepackage{fancyhdr}
 
\theoremstyle{plain}
\newtheorem{theorem}{Theorem}[section]

\begin{document}

\providecommand{\tr}{\ensuremath{\mathrm{tr}}}
\providecommand{\supp}{\ensuremath{\mathrm{supp}\ }}
\providecommand{\ad}{\ensuremath{\mathrm{ad}\ }}

\providecommand{\Ks}{\ensuremath{\mathcal{K}}}

\providecommand{\Hs}{\ensuremath{\mathcal{H}}}
\providecommand{\B}{\ensuremath{\mathcal{B}}}

\providecommand{\Ss}{\ensuremath{\mathcal{S}}}
\providecommand{\C}{\ensuremath{\mathbb{C}}}
\providecommand{\R}{\ensuremath{\mathbb{R}}}
\providecommand{\Z}{\ensuremath{\mathbb{Z}}}
\providecommand{\N}{\ensuremath{\mathbb{N}}}
\providecommand{\1}{\ensuremath{\mathbbm{1}}}

\providecommand{\T}{\ensuremath{\boldsymbol{T}}}
\providecommand{\A}{\ensuremath{\mathfrak{A}}}
\providecommand{\W}{\ensuremath{\mathfrak{W}}}
\providecommand{\xv}{\ensuremath{\mathbf{x}}}
\providecommand{\kv}{\ensuremath{\mathbf{k}}}
\providecommand{\dif}{\ensuremath{\mathrm{d}}}

\providecommand{\tensor}{\ensuremath{\otimes}}

\providecommand{\com}[1]{{\bf \ $\star $ #1 $\star$\ }}

\providecommand{\betrag}[1]{\ensuremath{\left| #1 \right|}}
\providecommand{\norm}[1]{\ensuremath{\left\| #1 \right\|}}
\providecommand{\wick}[1]{\ensuremath{\text{\bf :} #1 \text{\bf :}}}
\providecommand{\bra}[1]{\ensuremath{\left\langle {#1} \right|}}
\providecommand{\ket}[1]{\ensuremath{\left|{#1} \right\rangle}}


\title{Averaged Energy Inequalities for the Non-Minimally Coupled Classical Scalar Field}

\author{Christopher J. Fewster}
\email[email: ]{cjf3@york.ac.uk}
\author{Lutz W. Osterbrink}
\email[email: ]{lwo500@york.ac.uk}
\affiliation{Department of Mathematics, University of York, Heslington, York YO10 5DD, United Kingdom}
\date{\today}

\begin{abstract}
The stress energy tensor for the classical non-minimally coupled scalar field is known not to satisfy the point-wise energy
conditions of general relativity. 
In this paper we show, however, that local averages of the classical stress energy tensor satisfy certain inequalities. 
We give bounds for averages along causal geodesics and show, e.g., that in Ricci-flat background spacetimes, ANEC and AWEC are satisfied. 
Furthermore we use our result to show that in the classical situation we have an analogue to the phenomenon of {\it quantum interest}.
These results lay the foundations for analogous energy inequalities for the quantised non-minimally coupled
fields, which will be discussed elsewhere.
\end{abstract}

\maketitle
 
 
\section{Introduction}
Although the classical point-wise energy conditions are the cornerstone of many important results in classical general
relativity (e.g., the singularity theorems), it is well known that some classical matter models can violate them.
A standard example is the non-minimally coupled classical scalar field $\phi$, with field equation (see appendix \ref{convs} for sign conventions)
\begin{equation}\label{eq_weq}
(\square_{g}+m^{2}+\xi R)\phi =0,
\end{equation}
where $R$ is the Ricci scalar curvature and $\square_{g}$ is the d'Alembertian with respect to the metric~$g$ on the spacetime $M$. Since we are considering the classical field, the interpretation of $m$ is an inverse characteristic ``Compton wavelength''.

That the energy density for the non-minimal coupling can violate the weak energy condition can easily be seen.
Following the simple example given in~\cite{FR01}, we assume a solution~$\phi$ of the wave equation in Minkowski spacetime
with $m=0$ and $\xi>0$, propagating in one spatial direction, say the $x$-direction. Then any
$\mathcal{C}^2(\mathbb{R})$ function $h(u)$ defines a solution $\phi (t,x)=h(t-x)$ to the wave equation (\ref{eq_weq}).
Let $h'$ be the derivative of $h$, so that $\partial_{t}\phi=-\partial_{x}\phi=h'(t-x) $. In this
situation, the energy density component $T_{tt}$ of the stress-energy tensor, whose general expression (\ref{eq_st0})
is given in the next section, reduces at the point~$(t,x)$, to the form
\begin{equation}\label{eq_ed_example1}
T_{tt}=(1-2\xi)( h ')^{2}-2\xi h '' h, 
\end{equation}
with $h$ and its derivatives evaluated at $t-x$.
Since $h$ is allowed to be an arbitrary twice differentiable function, we can choose it to be positive (negative) and to have a local minimum (maximum) at the origin, i.e. $h$ and $h ''$ have the same sign and $ h '$ vanishes. Then $\phi$ obviously has a negative energy density $T_{tt}$ on the hyperplane $t=x$, since the non-vanishing part in~(\ref{eq_ed_example1}) is strictly negative.
Analogous examples can be found for more general situations, e.g., where $m> 0$ or $\xi<0$ or where the spacetime is not flat. With the same arguments as used above, one can find that the point-wise null energy condition is also violated.

In this paper, we will show that there are, nonetheless, constraints on
local averages of the stress-energy tensor for the non-minimally coupled
scalar field. We take our inspiration from quantum field theory, where
violations of the energy conditions are in fact inevitable~\cite{EGJ65}.
For example, the {\em minimally} coupled scalar field respects the weak energy
condition, but its quantisation admits states with negative energy densities. 
However, quantum field theory appears to contain mechanisms (related to
the uncertainty principle) which limit
the magnitude and duration of energy condition violation. These
mechanisms are expressed in so-called Quantum Energy Inequalities~(QEIs) (see~\cite{R04,F03}) which give state-independent lower bounds on certain weighted
averages of the energy density, using smooth compactly supported
weights. Applying the same basic idea to the non-minimally coupled classical scalar
field, we obtain lower bounds which are typically controlled by the geometry and the absolute value of
the field in the region of interest. Importantly, the 
derivatives of the field do not appear in the lower bound,
and this allows us to infer that large, long-lasting violations of the
energy conditions must be associated with large amplitude field
configurations or large curvature. 

A further inspiration for our approach stems from an argument 
presented in Sec.~4.3 of~\cite{HE73} in
connection with violations of the strong energy condition by the
minimally coupled classical field. There, the (unweighted) integral of
$T_{ab}W^aW^b-\frac{1}{2}W^aW_a T^{b}_{\phantom{b}b}$ over
a spacetime volume $\mathcal{U}$ (with $W^a$ a smooth timelike unit
vector field) was shown to be equal to a positive term plus an
integral over the boundary $\partial\mathcal{U}$, which was argued to be
small. By using averages with smooth compactly supported weight, our approach avoids the introduction
of a boundary term and leads to rigorous lower bounds. Furthermore, our
results lay the foundations for QEIs on the quantized non-minimally coupled scalar
field, which we will discuss elsewhere.

\section{Averaging on a causal geodesic}
\subsection{Main Result}
The stress energy tensor for the non-minimally coupled scalar field, and the wave equation~(\ref{eq_weq}), can be derived from its Lagrangian
\begin{equation}\label{eq_lag}
L=\frac{1}{2}\left\{ (\nabla \phi)^2 -(m^2+\xi R)\phi^2 \right\}.
\end{equation} 
Variation of the action with respect to $g^{\mu\nu}$ leads to the expression for the stress energy tensor, which is given by
\begin{equation}\label{eq_st0}
T_{\mu \nu}=\left(\nabla_\mu \phi\right)\left( \nabla_\nu \phi\right) +\frac{1}{2}g_{\mu\nu} \left(m^2\phi^{2}-(\nabla \phi)^2\right)+\xi \left\{ g_{\mu\nu}\square_{g} -\nabla_\mu \nabla_\nu-G_{\mu\nu}\right\} \phi^2,
\end{equation}
where $G_{\mu\nu}$ is the Einstein tensor.
One can easily see that this expression is consistent with that for the minimally coupled scalar field.
Furthermore, this expression reduces ``on shell'', i.e., for a $\mathcal{C}^2(M)$ solution of (\ref{eq_weq}), to
\begin{eqnarray}\label{eq_st1}
T_{\mu\nu}&=&(1-2\xi)\left(\nabla_\mu\phi\right)\left(\nabla_\nu\phi\right)+\frac{1}{2}\left(1-4\xi\right)g_{\mu\nu}\left(m^2\phi^2-(\nabla \phi)^2\right)
\nonumber\\
&&{}-\xi\left(2\phi\nabla_\mu\nabla_\nu \phi+R_{\mu\nu}\phi^2\right)+\frac{1}{2}\left(1-4\xi\right)g_{\mu\nu}\xi R\phi^2.
\end{eqnarray}
Even though the field equation (\ref{eq_weq}) and the Lagrangian (\ref{eq_lag}) for non-minimal coupling in Ricci-flat spacetimes reduce to those of minimal coupling, the stress energy tensor does not.
The reason is that the variational derivative defining $T_{\mu\nu}$ involves varying the action over non-flat metrics as well as flat ones. 

Now let $\gamma $ be a causal geodesic with affine parameter $\lambda$. We
will be interested in expressions of the form
\begin{equation}\label{eqn_geodaver}
\int_{\gamma} \mathrm{d} \lambda \ T_{\mu\nu}u^\mu u^\nu,
\end{equation}
where $u$ is a vector field with compact support on $\gamma$. For our purposes, we restrict to situations where $u$
is  a $\mathcal{C}^2_0(TM)$ vector-field, tangent to $\gamma$, i.e., we can always find a real-valued function
 $f\in \mathcal{C}^{2}_{0}(\mathbb{R})$ such that  $u=f\dot{\gamma}$. This, together with the fact that $\gamma $ is an affinely parametrized geodesic, gives $u^\mu u^\nu\nabla_\mu\nabla_\nu \phi =f^2 \partial^2_\lambda\phi$ .
Therefore, inserting expression~(\ref{eq_st1}) into~(\ref{eqn_geodaver}), we get
\begin{eqnarray}\label{eqn_set_geodav}
\int_{\gamma} \mathrm{d} \lambda \ T_{\mu\nu}u^\mu u^\nu
&=&\frac{1}{2}\int_{\gamma} \mathrm{d} \lambda \ (\partial_\lambda \phi)^2f^2+\frac{1}{2}\left(1-4\xi\right)\int_{\gamma} \mathrm{d} \lambda \ \phi^2f^2\left\{m^2\phi^2+h^{\mu\nu}(\nabla_\mu \phi)(\nabla_\nu \phi)\right\}
\nonumber\\
&&{}-2\xi  \int_{\gamma} \mathrm{d} \lambda \ f^2\phi \partial^2_\lambda \phi-\xi\int_{\gamma} \mathrm{d} \lambda \ \phi^2 f^2 \left\{  R_{\mu\nu}\dot{\gamma}^\mu \dot{\gamma}^\nu-\frac{1}{2}\left(1-4\xi\right)\dot{\gamma}^2 R\right\}
\end{eqnarray}
where we introduced $h^{\mu\nu}=u^{\mu}u^{\nu}-u^{2}g^{\mu\nu}$, which is positive semidefinite on $TM$
because  $u$ is non-space-like, as $\gamma$ is causal. Under the additional assumption that $\xi\leq 1/4$,
the first two terms on the right hand side of (\ref{eqn_set_geodav}) are positive. Neglecting the curvature terms, there is then only one term left, which can be either positive or negative, depending on the value of the field and its second derivative. 
But it is possible to write this term as a difference of positive terms. To do so, we use a simple identity, which was used in \cite{EFV05} to derive energy inequalities in quantum mechanics. In a slightly different form, it is given by
\begin{equation}\label{eqn_geodident}
2f^2\phi\partial^2_\lambda\phi+\partial_\lambda\left\{\phi\partial_\lambda (f^2\phi)-f^2\phi\partial_\lambda\phi\right\}=2\partial_\lambda \left\{f\phi \partial_\lambda (f\phi)\right\}-2[\partial_\lambda (f\phi)]^2+2\phi^2(\partial_\lambda f)^2,
\end{equation}
where we understand that $\phi=\phi \circ \gamma(\lambda)$. Its proof is a straightforward
calculation that, in a more general form, we give in appendix \ref{app_identity}.
Now since $f$ is a function of compact support, we can integrate (\ref{eqn_geodident}) and get (after multiplying by $-\xi$)
\begin{equation}\label{eqn_geodident2}
-2\xi \int_{\gamma} \mathrm{d} \lambda \ f^2\phi \partial^2_\lambda\phi=2\xi\int_{\gamma} \mathrm{d} \lambda \ [\partial_\lambda (f\phi)]^2-2\xi\int_{\gamma} \mathrm{d} \lambda \ \phi^2(\partial_\lambda f)^2.
\end{equation}
The expression on the right hand side is obviously a difference of positive terms and if $\xi$ is not
negative,  the first term on the right hand side is non negative and we obtain the following result by
putting (\ref{eqn_geodident2}) into (\ref{eqn_set_geodav}).
\begin{theorem}\label{thm_classgeod}
Let $\gamma$ be a causal geodesic with affine parameter $\lambda$ in a spacetime $(M,g)$.
Furthermore, let $T_{\mu\nu}$ be the stress-energy tensor of the non-minimally coupled classical scalar field with
coupling constant $\xi\in [0,1/4]$. For every
 real-valued function $f\in \mathcal{C}^{2}_{0}(\mathbb{R})$ the inequality
\begin{equation}\label{eqn_geodineq}
\int_{\gamma} \mathrm{d} \lambda \ T_{\mu\nu} \dot{\gamma}^\mu \dot{\gamma}^\nu f^2
\geq -2\xi\int_{\gamma} \mathrm{d} \lambda \ \left\{(\partial_{\lambda} f)^{2}+\frac{1}{2} R_{\mu\nu}\dot{\gamma}^\mu \dot{\gamma}^\nu f^2 -(\frac{1}{4}-\xi)R \dot{\gamma}^2f^{2}\right\}\phi^{2} 
\end{equation}
is satisfied ``on-shell''.
\end{theorem}
In particular this result includes the case of conformal coupling, i.e., where $\xi=\xi_c$ with
\begin{equation}
\xi_c=\frac{1}{4}\frac{n-2}{n-1}.
\end{equation}
One can also check that both sides of (\ref{eqn_geodineq}) are invariant under reparametrisation of the affine parameter $\lambda\rightarrow \tilde\lambda=\alpha \lambda+\beta$, if one also makes the replacement
\begin{equation}
f(\lambda )\rightarrow \tilde f(\tilde \lambda)=\sqrt{\alpha}\ f([\tilde \lambda -\beta]/\alpha).
\end{equation}
Now consider the case where $(M,g)$ is a vacuum solution to the Einstein equation, with vanishing cosmological constant, so the curvature terms in (\ref{eqn_geodineq}) vanish.
As we are interested in $\mathcal{C}^2$ solutions $\phi$, the maximal amplitude
\begin{equation}\label{eq_phimax}
\phi_{max} [\phi,\Omega]=\sup_{p\in \Omega}|\phi (p)|
\end{equation}
is finite for all compact regions $\Omega$ of $M$.
In any case we may bound $\phi$ by $\phi_{max}[\phi,\supp f]$ in the remaining contribution to (\ref{eqn_geodineq}), thus obtaining 
\begin{equation}\label{eq_averphimax}
\int_{\gamma} \mathrm{d} \lambda \ T_{\mu\nu} \dot{\gamma}^\mu \dot{\gamma}^\nu f^2
\geq -2\xi\phi^{2}_{max} [\phi,\supp f] \int_{\gamma}\mathrm{d} \lambda \ (\partial_{\lambda} f)^{2},
\end{equation}
which can be used to analyze local averages of, e.g. the energy density in a quite general form. Due to (\ref{eq_averphimax}), it is immediately obvious that, for a fixed coupling constant, the extent of energy condition violation is controlled by the maximal field amplitude. 

Inequalities of the form (\ref{eq_averphimax}) can also be derived for more general spacetimes. This usually involves a loss of generality, such as restricting the class of geodesics considered or the value of the coupling constant $\xi$. For example in a vacuum spacetime with $n\geq 3$ and cosmological constant $\Lambda$ one has $R_{\mu\nu}=2g_{\mu\nu}\Lambda/(n-2)$. Therefore the curvature dependent terms on the right hand side in (\ref{eqn_geodineq}) are 
\begin{equation}\label{eqn_curvtermcosm}
-2\xi\int_{\gamma} \mathrm{d} \lambda \ \left\{\frac{1}{2} R_{\mu\nu}\dot{\gamma}^\mu \dot{\gamma}^\nu f^2 -(\frac{1}{4}-\xi)R \dot{\gamma}^2f^{2}\right\}\phi^{2}=\xi\left(1-\xi\frac{4n}{n-2}\right)\Lambda \int_{\gamma} \mathrm{d} \lambda \  \dot{\gamma}^2f^{2}\phi^{2}.
\end{equation}
For a light-like geodesic $\gamma$, this term vanishes and if $\phi_{max}$ is defined, we obtain (\ref{eq_averphimax}). This result is independent of the sign of the cosmological constant.
However, if the geodesic is timelike, (\ref{eqn_curvtermcosm}) is non-negative either for $\Lambda\geq 0$ and $\xi\in[0,\xi_\Lambda]$ or for $\Lambda\leq0$ and $\xi\in [\xi_\Lambda,1/4]$, where 
\begin{equation}
\xi_\Lambda=\frac{n-2}{4n}.
\end{equation}
Under either circumstance, one can omit the term (\ref{eqn_curvtermcosm}) in (\ref{eqn_geodineq}). Again, existence of $\phi_{max}$ leads to the lower bound (\ref{eq_averphimax}). Note that $0<\xi_\Lambda<\xi_c<1/4$ for any spacetime dimension $n>2$.

\subsection{Scaling arguments}\label{geod_scal}
We now investigate how the lower bound in
(\ref{eq_averphimax}) changes under rescaling of a fixed smearing function $f$.  
For this purpose we introduce the function 
\begin{equation}\label{eqn_scalfctn}
f_{\lambda_{0}}(\lambda)=\lambda_{0}^{-1/2}\ f(\lambda /\lambda_{0}),\qquad \lambda_0>0,
\end{equation}
which is chosen in such a way that the normalisation of $f_{\lambda_0}$ is independent of the choice of~$\lambda_0$, i.e.,
\begin{equation}
\int   \mathrm{d}\lambda \ f^2_{\lambda_0}(\lambda)=\int   \mathrm{d}\lambda \ f^2 (\lambda)=1,\qquad \forall\  \lambda_0>0.
\end{equation}
Furthermore, we can define  
\begin{equation}
C_f=\int  \mathrm{d}\lambda \ (\partial_{\lambda} f)^{2},
\end{equation}
which again is a constant since $f$ is fixed.
Now let us suppose that the field $\phi$ is bounded on a complete geodesic $\gamma$. Then averaging with respect to $f_{\lambda_{0}}$ gives
\begin{equation}
\int_{\gamma} \  \mathrm{d}\lambda \ T_{\mu\nu}\dot{\gamma}^{\mu}\dot{\gamma}^{\nu} \
f^{2}_{\lambda_{0}}(\lambda)\geq  -\frac{2\xi C_f}{\lambda^{2}_{0}}\phi^{2}_{max} [\phi,\gamma]\ .
\end{equation}
In the limit $\lambda_0\to 0$, we obtain consistency with the fact that
the pointwise energy conditions can be arbitrarily badly violated; on
the other hand, in the scaling limit where $\lambda_{0}$ tends to infinity, 
we find \footnote{Actually this result would still be true if the field
$\phi$ on the geodesic obeys $\phi(\lambda)/\lambda\to 0$ as~$|\lambda|\to\infty$.}
\begin{equation}\label{eq_liminf}
\liminf_{\lambda_{0}\rightarrow \infty}  \int_{\gamma} \  \mathrm{d}\lambda \ T_{\mu\nu}\dot{\gamma}^{\mu}\dot{\gamma}^{\nu} \ f^{2}_{\lambda_{0}}(\lambda)\geq 0.
\end{equation}
If $T_{\mu\nu}\dot{\gamma}^{\mu}\dot{\gamma}^{\nu}$ is integrable on $\gamma$, we can apply the dominated convergence theorem and get the following:
\begin{theorem}
Let $\gamma$ be a complete causal affinely parametrized geodesic in a Ricci-flat spacetime $(M,g)$. Let $T_{\mu\nu}$ be the stress energy tensor for the non-minimally coupled scalar field with coupling constant $\xi\in [0,1/4]$. If the field is bounded on the geodesic $\gamma$, then we have
\begin{equation}\label{eq_averint}
\int_{\gamma} \  \mathrm{d}\lambda \ T_{\mu\nu}\dot{\gamma}^{\mu}\dot{\gamma}^{\nu}\geq 0,
\end{equation}
if the expression on the left-hand side exists.
\end{theorem}
We see that in the situation where $\gamma$ is time-like (light-like), 
the expression (\ref{eq_averint}) reduces to the AWEC (ANEC).
The inequality (\ref{eq_liminf}) may be regarded as a generalisation of these conditions, see e.g.,
\cite{Y95}. In Ricci-flat spacetimes it is easy to give a direct proof
of these conditions under the hypothesis that $\phi\,\partial_\lambda\phi\to 0$
on $\gamma$ at infinity.
See, e.g., Sec.~II of \cite{K91}, where it is noted in
passing for ANEC in Minkowski space (the main focus being the quantum case). Our
result achieves this end with weaker hypotheses.

A slight variant on the above approach is to estimate $T_{\mu\nu}\dot{\gamma}^{\mu}\dot{\gamma}^{\nu}$
in (\ref{eq_averphimax}) by its supremum over some open interval $I$. Then
for every normalised smooth $f$ compactly supported in $I$, we obtain a bound
\begin{equation}
\sup_{I} T_{\mu\nu}\dot{\gamma}^{\mu}\dot{\gamma}^{\nu} 
\geq  -2\xi C_f\phi^{2}_{max} [\phi,I]\ .
\end{equation}
It is well
known (but see e.g., \cite{FT98} for details), that the infimum of $C_f$ over
functions of this class is $\inf_f C_f=\pi^2 / \Gamma^2$ where $\Gamma$
is the length of the interval $I$. As we are free to optimise the
right-hand side over $f$, we obtain
\begin{equation}
\sup_{I} T_{\mu\nu}\dot{\gamma}^{\mu}\dot{\gamma}^{\nu} \geq 
-\frac{2\xi \pi^2}{\Gamma^2}\phi^{2}_{max} [\phi,I]\ ,
\end{equation}
showing that long-lasting negative energy densities of large magnitude must be associated
with large magnitudes of the field.

\subsection{Energy interest}\label{geod_energ}
The previous results suggest that there should be an
analogue to the phenomenon in quantum field theory and quantum mechanics known as {\it quantum interest}~\cite{FR99}.
Roughly speaking, negative energy densities occurring in these theories have to be overcompensated,
i.e., a negative energy density pulse has to be accompanied by an even larger positive pulse, so that the
overall averaged energy density is positive. Additionally there are restrictions on
the amplitude and time separation
for these pulses. For further reading see \cite{FT98} and references therein.
We will discuss this phenomenon in the classical situation for the non-minimally coupled scalar field.
Consider a complete time-like geodesic $\gamma$ in a Ricci-flat spacetime,
and, as an illustration, suppose that the energy density takes the form
\begin{equation}\label{eq_enerdens}
T_{\mu\nu}\dot{\gamma}^{\mu}\dot{\gamma}^{\nu}=[-\delta (\lambda)+(1+\varepsilon)\ \delta (T-\lambda)] \rho_{0},
\end{equation}
for some positive constants $T$ and $\rho_0$.
(Of course this should be regarded as an idealised model of a smooth,
highly peaked configuration.) We are interested in what expression (\ref{eq_averphimax}) can tell us
about the parameters $\rho_{0}$, $T$ and $\varepsilon$, in relation to 
$\phi_{max}=\phi_{max}[\phi,\gamma]$, which we assume to be finite.  Now it is clear that the unweighted average energy density on $\gamma$ is given by $\varepsilon\rho_{0}$. 
Using (\ref{eq_enerdens}) in (\ref{eq_averphimax}) and integrating by parts once gives
\begin{equation}\label{eq_posint}
\int \mathrm{d} \lambda \ f\left\{-2\xi\phi^{2}_{max}  \ \partial^{2}_{\lambda} f +[-\delta (\lambda)+(1+\varepsilon)\ \delta (\lambda -T)] \rho_{0} f \right\}\geq 0,
\end{equation}
where we assume $f$ to be a real-valued function in $ \mathcal{C}^\infty_0(\mathbb{R})$.
An implication of (\ref{eq_posint}) is that there exist no square-integrable solutions to the eigenvalue problem (see \cite{FT98}) 
\begin{equation}\label{eq_eigvalprob}
-f'' (x)+[-\alpha \delta (x) +\beta \delta (x-T) ]f(x)=-k^{2}f(x),
\end{equation}
for $k>0$, where the parameters are given by
\begin{equation}\label{eqn_params}
\alpha=\frac{\rho_{0}}{2\xi \phi^{2}_{max} } \qquad\textrm{and}\qquad \beta=(1+\varepsilon)\ \frac{\rho_{0}}{2\xi \phi^{2}_{max}  }.
\end{equation}
That is, whenever this eigenvalue problem has a solution, we are in contradiction with the result of theorem \ref{thm_classgeod}. It can be shown \cite{FT98} that solutions to (\ref{eq_eigvalprob}) do not exist if and only if the parameters (\ref{eqn_params}) satisfy
\begin{equation}
0\leq  \alpha T <1\qquad \textrm{ and }\qquad \alpha T \leq \beta T(1-\alpha T).
\end{equation}
Therefore theorem \ref{thm_classgeod} tells us that
\begin{equation}\label{eqn_interest}
0\leq  \rho_{0}T <2\xi \phi^{2}_{max}\qquad \textrm{ and }\qquad \varepsilon \geq \frac{\rho_{0}T}{2\xi\phi^{2}_{max}-\rho_{0} T}.
\end{equation}
These inequalities tell us two things, the first of which is that there is a restriction for time separation
and  amount of energy density $\rho_{0}$, in terms of the maximal field amplitude and the coupling constant. The second result is that in every non-trivial situation, the amount of positive energy overcompensates the negative energy density since $\varepsilon>0$.

This phenomenon suggests that non-minimally coupled scalar fields cannot be used to produce long lasting
violations of the second law of thermodynamics, except by very large values of the field amplitude.
See \cite{F78} for related remarks in the context of quantum field theory (which actually led to
the development of the quantum energy inequalities mentioned in the introduction).

\section{Averaging over a spacetime volume}
We now show that it is also possible to find lower bounds similar to those in \ref{thm_classgeod} for volume averaging.
For this purpose we take a compactly supported nowhere space-like vector field $u\in \mathcal{C}^{\infty}_{0}(TM)$. Local averages 
on $\supp u$ are then given by expressions of the form
\begin{equation}\label{eq_avst1}
\int \mathrm{d} vol_{g} \ T_{\mu\nu}u^\mu u^\nu.
\end{equation}
We want to find a similar procedure to decompose the averaged stress energy tensor, as done in the previous section.
For this purpose, let us look at the first term in the second row of~(\ref{eq_st1}), i.e. $2\phi\nabla_\mu\nabla_\nu \phi+R_{\mu\nu}\phi^2$.
Contracted with $u$, this term can be reformulated using the identity 
\begin{eqnarray}\label{eq_ident}
\lefteqn{2\phi u^{\mu}u^{\nu}\nabla_{\mu}\nabla_{\nu}\phi +R_{ \mu \nu}u^{\mu}u^{\nu}\phi^{2}-\nabla_\mu [\phi u^\mu u^\nu \nabla_\nu \phi-\phi \nabla_\nu \left(u^\mu u^\nu \phi\right)]}\nonumber\\
&=& \left\{(\nabla \cdot u)^{2} + \tr (\nabla u)^{2}\right\}\phi^{2} -2\left\{\nabla \cdot \left(\phi u\right)\right\}^2+2\nabla_\mu [u\phi \nabla_\nu \phi u^\nu ],
\end{eqnarray}
where    
\begin{equation}
\tr (\nabla u)^{2}=\left(\nabla_{\mu}u^{\nu}\right)\left(\nabla_{\nu}u^{\mu}\right).
\end{equation}
The expression (\ref{eq_ident}) can be proved by a straightforward, but lengthy, calculation, which we give in appendix \ref{app_identity}.

Integrating both sides of (\ref{eq_ident}) yields the identity
\begin{eqnarray}\label{eq_intident}
\lefteqn{-\xi\int \mathrm{d} vol_{g} \  u^{\mu}u^{\nu}\left(2\ \phi\nabla_\mu\nabla_\nu \phi+R_{\mu\nu}\phi^2\right)}\nonumber\\
&&=2\xi\int  \mathrm{d} vol_{g} \  \left\{\nabla \cdot (u\phi)\right\}^{2}-\xi\int \mathrm{d} vol_{g}\ [(\nabla \cdot u)^{2} + \tr (\nabla u)^{2}]\phi^{2},
\end{eqnarray}
since, due to $u$ being compactly supported, the divergence terms have vanishing integral. This identity is the generalisation of (\ref{eqn_geodident2}) for volume averages.

Returning to the averaged stress energy tensor and inserting the ``on shell'' stress energy tensor (\ref{eq_st1}) into expression (\ref{eq_avst1}), we find
\begin{eqnarray}\label{eq_avst2}
\lefteqn{\int \mathrm{d} vol_{g} \ T_{\mu\nu}u^\mu u^\nu}\nonumber\\
&=&\frac{1}{2}\int \mathrm{d} vol_{g} \ (u^{\mu}\nabla_\mu \phi)^2+\frac{1}{2}(1-4\xi)\int  \mathrm{d} vol_{g} \ \left[m^2u^2 \phi^2+ h^{\mu\nu}\left(\nabla_\mu \phi\right)\left(\nabla_\nu \phi\right)\right] \nonumber\\
&&-\xi \int  \mathrm{d} vol_{g} \  \left\{u^{\mu}u^{\nu}\left(2\ \phi\nabla_\mu\nabla_\nu \phi+R_{\mu\nu}\phi^2\right)-\frac{1}{2}(1-4\xi)Ru^{2} \phi^2 \right\}.
\end{eqnarray}
Here again we used $h^{\mu\nu}=u^{\mu}u^{\nu}-u^{2}g^{\mu\nu}$, which is positive semidefinite on $TM$ because $u$ is non-space-like.
Therefore the expression in the middle line of (\ref{eq_avst2}) is positive if $\xi\leq 1/4$.

Using the identity (\ref{eq_intident}), we see that the averaged stress energy tensor can be written in the form
\begin{eqnarray}\label{eq_avst3}
\lefteqn{\int \mathrm{d} vol_{g} \ T_{\mu\nu}u^\mu u^\nu}\nonumber\\
&=&\frac{1}{2}\int \mathrm{d} vol_{g} \ (u^{\mu}\nabla_\mu \phi)^2+\frac{1}{2}(1-4\xi)\int \mathrm{d} vol_{g} \ \left[m^2u^2 \phi^2+ h^{\mu\nu}\left(\nabla_\mu \phi\right)\left(\nabla_\nu \phi\right)\right]\nonumber\\
&&+2\xi\int  \mathrm{d} vol_{g} \  [\nabla \cdot (u\phi)]^{2}-\xi\int \mathrm{d} vol_{g}\ [(\nabla \cdot u)^{2} + tr (\nabla u)^{2}-\frac{1}{2}(1-4\xi)R u^2]\phi^{2}  .
\end{eqnarray}
Here we see that, in addition to the non-negative terms mentioned above,
the first term in the bottom line of (\ref{eq_avst3}) is also
non-negative for  $\xi\geq 0$. This observation proves the following result
\begin{theorem}\label{thm_class}
Let $T_{\mu\nu}$ be the stress-energy tensor of the non-minimally coupled classical scalar field, with
coupling constant $\xi\in [0,1/4]$ on a spacetime $M$. For any nowhere space-like vector field $u\in \mathcal{C}^{2}_{0}(TM)$, the inequality
\begin{equation}\label{eqn_thmvolint}
\int \mathrm{d} vol_{g} \ T_{\mu\nu}u^\mu u^\nu 
\geq -2\xi\int \mathrm{d} vol_{g} \ \left\{\frac{1}{2}(\nabla \cdot u)^{2} + \frac{1}{2} \tr (\nabla u)^{2}-(\frac{1}{4}-\xi)R u^2\right\}\phi^{2}
\end{equation}
is satisfied ``on-shell''.
\end{theorem}
Thus, there exist lower bounds for the volume averages of the stress-energy density of a form similar to (\ref{eqn_geodineq}).
We will not discuss applications of theorem \ref{thm_class} here, but remark that the lower bound is again controlled by $\phi$ and not its derivatives. The form of (\ref{eqn_thmvolint}) suggests, that one can expect results similar to those in \ref{geod_scal} and \ref{geod_energ}.

\section{Conclusion}
We have shown that there exist lower bounds for certain averages of the stress energy
tensor for the classical non-minimally coupled scalar field. 
These show similarities with Quantum Energy Inequalities and entail that large,
long-lasting violations of the energy conditions are
associated with large magnitude field configurations or large
spacetime curvature (or coupling constants outside the range $[0,1/4]$). As corollaries, we showed that the non-minimally coupled classical scalar field on a Ricci-flat
spacetime always satisfies the AWEC and ANEC condition, provided $\phi$ is bounded on the
geodesic in question.

Furthermore, we showed that there exists an energy interest phenomenon,
i.e., a pulse of negative energy density is always accompanied by an overcompensating
positive one. The same analysis showed that there are also restrictions on the amplitudes
and time separation of these pulses. These effects are well known for the minimally
coupled scalar {\it quantum} field. 
 
It is worth mentioning that theorems \ref{thm_classgeod} and \ref{thm_class} can also be generalised
to potentials other than the mass term. If the vector field $u$ is light-like, any potential is possible. If we average with respect to a time-like vector-field $u$, one can show that the results are still true for any potentials $V[\phi]$, replacing $\frac{1}{2} m^{2}\phi^{2}$ in the Lagrangian, that satisfy 
\begin{equation}\label{eq_const1}
V[\phi]-2\xi \phi \frac{\delta V}{\delta \phi} [\phi]\geq 0,
\end{equation} 
for some $\xi\in [0,1/4]$. This relation originates in the fact that we used the wave equation of the field in (\ref{eq_st1}). As an example, such a potential is
$V[\phi]=\mu^{2}\phi^{2k}$, for positive constants $\mu$ and $k$, with $\xi k\leq 1/4$. It is clear that for minimal coupling, the condition (\ref{eq_const1}) reduces to the requirement that $V[\phi]$ is a positive potential.

\acknowledgments
The authors are grateful to Calvin Smith for a careful reading of the manuscript. 

\appendix
\section{Conventions}\label{convs}
The conventions used are~$(-,-,-)$ in the classification of Misner, Thorne and Wheeler~\cite{MTW73}.
These can be found in, e.g., Birrell and Davies' book~\cite{BD82}.
In detail this means that the signature of the metric tensor of the spacetime is $(+--\dots)$.
Furthermore the Riemann tensor is defined by\footnote{In terms of Christoffel symbols, this is equivalent to
$ \displaystyle
R_{\rho \sigma \alpha}^{\ \ \ \ \beta}=\Gamma^\beta_{\sigma\alpha ,\rho}-\Gamma^\beta_{\rho \alpha, \sigma}+\Gamma^\beta_{\rho \gamma}\Gamma^\gamma_{\sigma \alpha} - \Gamma^\beta_{\sigma \gamma}\Gamma^\gamma_{\rho \alpha}
$.}
\begin{eqnarray}\label{eq_riemtensdef}
\left(\nabla_{\mu}\nabla_{\nu}-\nabla_{\nu}\nabla_{\mu}\right)\omega^{\sigma}=R_{\mu\nu\alpha}^{\ \ \ \ \sigma} \omega^{ \alpha},
\end{eqnarray}
where $\nabla$ is the Levi-Civita connection and $\omega$ is a vector field on the manifold . Using (\ref{eq_riemtensdef}), we see that for the Ricci tensor $R_{\rho\alpha}=R_{\rho \beta\alpha}^{\ \ \ \ \beta}$, we get  
\begin{eqnarray}
\left(\nabla_{\mu}\nabla_{\nu}-\nabla_{\nu}\nabla_{\mu}\right)\omega^{\nu}=R_{\mu\nu}\ \omega^{\nu}.
\end{eqnarray}
In these conventions, the Einstein equations with cosmological constant $\Lambda$ are 
\begin{equation}
G_{\mu\nu}+\Lambda g_{\mu\nu}=-\kappa T_{\mu\nu},
\end{equation} 
where the Einstein-tensor is defined in the usual way as $G_{\mu\nu}=R_{\mu\nu}-\frac{1}{2}Rg_{\mu\nu}$ and the coupling constant is given by $\kappa=8\pi G/c^{4}$. We use units in which $c=1$.
 
\section{Proof of an Identity}\label{app_identity}
In this appendix we will prove the identity (\ref{eq_ident}).
We begin by noting the identities
\begin{equation}
\phi \nabla_{\nu}\nabla_{\mu}(u^{\mu}u^{\nu}\phi)=\phi u^{\mu} u^{\nu}\nabla_{\mu}\nabla_{\nu}\phi -\nabla_{\mu}\left[ \phi u^{\mu}u^{\nu}\nabla_{\nu}\phi-\phi \nabla_{\nu}(u^{\mu}u^{\nu}\phi) \right]
\end{equation} 
and
\begin{equation}
\phi u^{\mu}\nabla_{\mu}\nabla_{\nu}(u^{\nu}\phi)=\nabla_{\mu}\left[ u^{\mu}\phi\nabla_{\nu}(u^{\nu}\phi) \right] - \left[\nabla \cdot (u\phi) \right]^{2}\phi^{2}.
\end{equation}
Using these, it is obvious that (\ref{eq_ident}) is proved, if we show that
\begin{eqnarray}\label{eqn_identproof1}
\lefteqn{\phi u^{\mu}u^{\nu}\nabla_{\mu}\nabla_{\nu}\phi + \phi \nabla_{\nu}\nabla_{\mu}(u^{\mu}u^{\nu}\phi) +R_{ \mu \nu}u^{\mu}u^{\nu}\phi^{2}}\nonumber\\
&&=
 \left\{(\nabla \cdot u)^{2} + \tr (\nabla u)^{2}\right\}\phi^{2}+2\phi u^{\mu}\nabla_{\mu}\nabla_{\nu}(u^{\nu}\phi) .
\end{eqnarray}
To see this we calculate the left hand side of (\ref{eqn_identproof1}) first. Since 
$
u^{\mu}\nabla_{[\mu}\nabla_{\nu ]}u^{\nu}=\frac{1}{2}R_{\mu\nu}u^{\mu}u^{\nu}
$, we get
\begin{eqnarray}
\textrm{LHS}&=&\phi u^{\mu}u^{\nu}\nabla_{\mu}\nabla_{\nu}\phi + \phi \nabla_{\nu}\nabla_{\mu}(u^{\mu}u^{\nu}\phi) + \phi^2 u^{\mu}\nabla_{\mu}\nabla_{\nu }u^{\nu}-\phi^2 u^{\mu}\nabla_{\nu}\nabla_{\mu }u^{\nu}\nonumber\\
&=&\phi u^{\mu}u^{\nu}\nabla_{\mu}\nabla_{\nu}\phi + \phi \nabla_{\nu} \left[ u^\mu u^\nu \nabla_\mu \phi +u^\mu \phi \nabla_\mu u^\nu +u^\nu \phi \nabla_\mu u^\mu  \right] \nonumber\\
&&\hspace{6cm} + \phi^2 u^{\mu}\nabla_{\mu}\nabla_{\nu }u^{\nu}-\phi^2 u^{\mu}\nabla_{\nu}\nabla_{\mu }u^{\nu}\nonumber\\
&=& \phi (\nabla_\nu u^\mu) u^\nu \nabla_\mu \phi + \phi u^\mu (\nabla_\nu u^\nu) \nabla_\mu \phi +2\ \phi u^{\mu}u^{\nu}\nabla_{\mu}\nabla_{\nu}\phi\nonumber\\
&&+\phi (\nabla_\nu u^\mu) (\nabla_\mu u^\nu) \phi+\phi u^\mu (\nabla_\mu u^\nu)\nabla_\nu \phi+0\nonumber\\
&&+\phi (\nabla_\mu u^\mu) (\nabla_\nu u^\nu) \phi +\phi  (\nabla_\mu u^\mu) u^\nu (\nabla_\nu \phi) +2\ \phi^2 u^\nu\nabla_\nu \nabla_\mu u^\mu \nonumber\\
&=&\left\{(\nabla \cdot u)^{2} 
+ \tr (\nabla u)^{2}\right\}\phi^{2}\nonumber\\
 &&+2\left\{\phi (\nabla_\nu u^\mu) u^\nu \nabla_\mu \phi + \phi u^\mu (\nabla_\nu u^\nu) \nabla_\mu \phi +\phi u^{\mu}u^{\nu}\nabla_{\mu}\nabla_{\nu}\phi + \phi^2 u^\nu\nabla_\nu \nabla_\mu u^\mu \right\}.
\end{eqnarray}
In the last step we simply reordered and used
$
\left(\nabla_{\mu}u^{\nu}\right)\left(\nabla_{\nu}u^{\mu}\right)=\tr (\nabla u)^{2}
$. To prove (\ref{eqn_identproof1}) it remains to show that
\begin{equation}
\phi u^{\mu}\nabla_{\mu}\nabla_{\nu}(u^{\nu}\phi)=\phi (\nabla_\nu u^\mu) u^\nu \nabla_\mu \phi + \phi u^\mu (\nabla_\nu u^\nu) \nabla_\mu \phi +\phi u^{\mu}u^{\nu}\nabla_{\mu}\nabla_{\nu}\phi + \phi^2 u^\nu\nabla_\nu \nabla_\mu u^\mu,
\end{equation}
which follows by the Leibniz rule. This completes the proof of~(\ref{eqn_identproof1}) and therefore, as remarked above, the proof of (\ref{eq_ident}). 
In a simplified, but analogous way, one proves (\ref{eqn_geodident}).

\end{document}